\documentclass[aps,pra,twocolumn]{revtex4}% Physical Review A
\usepackage{graphicx}% Include figure files
\usepackage{dcolumn}% Align table columns on decimal point
\usepackage{bm}% bold math
\usepackage{amssymb}

\begin{document}
\preprint{APS/123-QED}

\title{Single atom detection of calcium isotopes by atom trap trace analysis}

\author{S. Hoekstra, A. K. Mollema, R. Morgenstern, H. W. Wilschut and R. Hoekstra}
\affiliation{Atomic Physics, KVI, Rijksuniversiteit Groningen, Zernikelaan 25, 9747 AA, Groningen, The
Netherlands}

\date{\today}% It is always \today, today,

\begin{abstract}
We demonstrate a combination of an isotopically purified atom beam and a magneto-optical trap which enables the
single atom detection of all stable isotopes of calcium (40, 42, 43, 44, 46 and 48). These isotopes range in
abundance from 96.9 \% ($^{40}$Ca) to 0.004 \% ($^{46}$Ca). The trap is loaded from an atomic beam which is
decelerated in a Zeeman slower and subsequently deflected over an angle of 30$^{\circ}$ by optical molasses. The
isotope selectivity of the Zeeman slower and the deflection stage is investigated experimentally and compared
with Monte Carlo simulations.
\end{abstract}

\maketitle

\section{\label{sec:level1}Introduction}
Trace analysis of long-lived isotopes has become an important tool in modern science. From medical science to
environmental research, from nuclear safety to archeology, the capability to detect low-abundance long-lived
isotopes has opened many research fields~\cite{60}. For the detection of trace elements Atom Trap Trace Analysis
(ATTA) is a promising and potentially very powerful new technique~\cite{223,213}. In a recent experiment one
million year old groundwater from the Nubian Aquifer (Egypt) has been dated using ATTA, by detecting very small
traces of $^{81}$Kr~\cite{215}. Because there are no interferences from other elements ATTA has the potential to
also detect the long-lived isotope $^{41}$Ca at its natural abundance level of $10^{-14}$~\cite{111}. This would
open the possibility to perform radio-calcium dating with $^{41}$Ca, which has a half-life of $10^{5}$
years~\cite{111,110}. Furthermore, $^{41}$Ca could be used as a tracer to directly monitor the bone loss and
retention rates of human subjects in both research and diagnosis of osteoporosis~\cite{109}. As an alternative
to Accelerator Mass Spectrometry (AMS), ATTA may provide a small-scale and cost-effective detection method of
rare isotopes, such as the example of $^{81}$Kr showed. Recently ATTA has been compared with the established
technique of Low Level Counting (LLC)\cite{217}. A competing compact laser-based method is Resonance Ionization
Mass Spectrometry (RIMS), which has been successfully developed in recent years \cite{312,313,315}. RIMS
combines the selectivity of laser spectrometry with the technique of mass separation. Recently it has been shown
that ATTA is already sufficiently sensitive to detect $^{41}$Ca in enriched ($10^{-8}$) calcium samples for
biomedical applications \cite{65}. We have set up an ATTA experiment with the ultimate goal of detecting
$^{41}$Ca at the natural abundance level.

ATTA experiments combine various optical techniques, each of which is isotope selective. The mechanism of
isotope selection is the repeated excitation (by a laser) of an optically accessible electronic transition in
the neutral atom. Because of the isotope shift the scattering force induced by light of a fixed frequency is
different for the different isotopes. It is the ratio between the natural linewidth and the isotope shift of the
pumping transition that determines the selectivity that can be reached between neighboring isotopes. For the
calcium isotopes the isotope shift is about five times the natural linewidth; therefore isotope selectivity in
laser pumping of calcium isotopes is possible. Reducing the Doppler broadening of the transition is an essential
ingredient for the isotope selectivity. Therefore, cold atoms are an ideal and necessary tool for
ultra-sensitive isotope detection. Samples of cold atoms can be obtained by laser cooling and trapping in a
magneto-optical trap (MOT)~\cite{103}, which is the central element of an ATTA experiment.

The final sensitivity that can be reached in an ATTA experiment is limited either by the background of $^{40}$Ca
atoms or by the loading rate of the trap. The background of $^{40}$Ca atoms can be reduced by improving the
isotope selectivity, the loading rate can be increased by improving the efficiency. In the ATTA measurements
reported on in \cite{65} the isotope selectivity was the limiting factor for the sensitivity. In this article
the isotope selectivity has been investigated in detail. This results in a setup in which the search for
$^{41}$Ca atoms is no longer limited by the isotope selectivity but by the loading efficiency. The loading
efficiency is limited at the moment by the laser power available for the experiment.

In section~\ref{sec:setup} of this article the various parts of our experimental setup will be presented. In
section~\ref{sec:results} three different loading schemes for the MOT are compared with the purpose of improving
the isotope selectivity. Monte Carlo simulations were made to help understand the isotope selectivity of the
Zeeman slower and the deflection stage. Finally, experimental data of the detection of single calcium atoms in
the MOT are presented.

\section{Experimental setup}\label{sec:setup}
\begin{figure}[htb]
\includegraphics[width=\columnwidth]{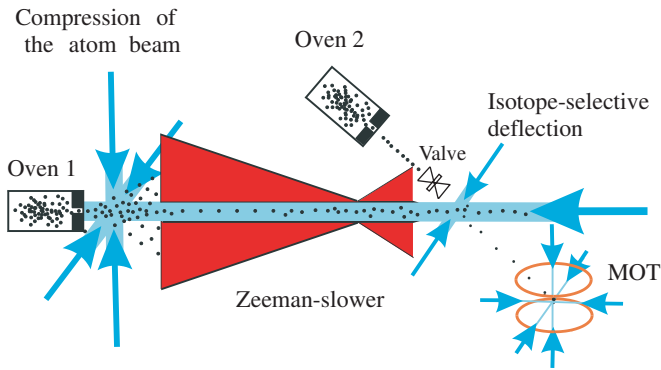}
\caption{\label{fig:setup}A schematic overview of the experimental setup.}
\end{figure}
Our experimental method is schematically depicted in figure~\ref{fig:setup}. The metallic calcium sample is
placed in oven 1 from which atoms are evaporated to create an atomic beam. The divergence of the atomic beam of
a selected isotope can be reduced using optical compression with laser beams perpendicular to the atomic
beam~\cite{249}. The other isotopes are not affected or even pushed out of the atomic beam. The second step in
the selection of the isotopes, and essential for the efficient loading of the magneto-optical trap, is a Zeeman
slower. It slows the atoms down by the absorption of photons from a counter-propagating laser beam~\cite{246}.
The Doppler shift resulting from the longitudinal velocity of the atoms is continuously compensated by a Zeeman
shift, induced by the magnetic field along the axis. The resulting slow atom beam leaving from the oven is
deflected over 30$^{\circ}$ in the direction of the Magneto-Optical Trap (MOT) by a standing wave field tilted
at the proper angle. This deflection stage is an essential step in the isotope selectivity and the total
sensitivity that can be reached. As will be shown in this paper only the selected isotope is effectively
deflected and guided into the trap. This improves the signal-to-noise ratio in the detection of the rare
isotopes, and thereby enables the use of a higher atom flux. The detection of the rare isotopes takes place in
the MOT. Only atoms of one isotope can be trapped for a given laser frequency. By scanning the laser frequency
and observing the fluorescence of the trapped atoms we detected all stable isotopes of calcium down to the
single atom level. The trap can also be loaded from a second oven; it is connected to the deflection chamber
with a valve as indicated in figure~\ref{fig:setup}. This second atom beam enables us to make a direct
comparison between the different methods by which the trap can be loaded. This will be presented in
section~\ref{sec:results}. The various parts of the experiment will be presented here in some more detail.

At 423 nm, calcium has a strong resonance transition from the ground $4s^{2}$ $^{1}$S$_{0}$ state to the $4s4p$
$^{1}$P$_{1}$ state which is rather well suited for laser cooling. The required 423 nm laser light for the
cooling is generated by frequency doubling of the output of an 846 nm diode laser (Toptica Photonics AG). By
amplification in a tapered amplifier up to 500 mW of 846 nm light is produced; after the frequency doubling with
an LBO crystal we typically have 55 mW of 423 nm light available for the locking of the laser (5 mW), the
compression stage (10 mW), the Zeeman slower (20 mW), the deflection stage (5 mW) and the MOT (15 mW). The
amount of laser power available is the limiting factor for the efficiency of the experiment, and could be
increased by investing in a more intense source of light at 846 nm. The main laser beam is split and frequency
shifted using beam-splitters and acousto-optical modulators (AOM). The laser frequency is locked to the cooling
transition of calcium by means of polarization spectroscopy~\cite{4}, which is done on an atomic beam from oven
2. The average trapping time of calcium in a MOT is limited to $\sim$ 20 ms due to a weak leak (10$^{-5}$) from
the $4s4p$ $^{1}$P$_{1}$ to the $^{1}$D$_{2}$ state. From the $^{1}$D$_{2}$ state roughly 75 \% decays to the
ground state within 3 ms, and can be recaptured if the diameters of the trapping laser beams are large enough.
The rest of the atoms are lost from the trap, and limit the trapping time. The trapping time can be increased by
repumping the atoms from the $^{1}$D$_{2}$ to the $5s$ $^{1}$P$_{1}$ state, from which they quickly decay back
to the ground state \cite{54}. The laser light required for the repumping is generated by a home built diode
laser operating at 672 nm, of which typically 5 mW is available for the MOT. Only for the measurements with
single atoms presented here the repump laser has been used.

The calcium atoms are placed in an oven from which the atoms are evaporated. The ovens have 10 exit channels
each with a diameter of 1 mm and a length of 10 mm. A ceramic tube around the oven holds tantalum wires which
are used to heat the oven to temperatures in the range of 400 to 600 $^{\circ}$C. The Maxwellian velocity
distribution for calcium atoms with a temperature of 600 $^{\circ}$C peaks at a velocity of 600 m/s.

The transverse velocity is less than one tenth of the longitudinal velocity, provided the mean free path of the
atoms is larger than the length of the exit channel of the oven. Directly after the atoms leave the oven we have
the possibility to apply transverse compression of the atom beam. The Doppler shift $\omega_{D}$ of the moving
atom, the isotope shift $I_{s}$ and the natural linewidth $\Gamma$ of the transition together determine the
effective scattering rate of a specific isotope. For a specific laser detuning $\delta$ and power $s_{0}$ the
scattering rate $\gamma_{p}$~\cite{205} is given by
\begin{eqnarray}
\gamma_{p}=\frac{s_{0}\Gamma/2}{1+s_{0}+(2(\delta +\omega _{D}+I_s{})/\Gamma )^{2}} \label{eqn:scattering}
\end{eqnarray}
As can be seen from table~\ref{tab:isotopes} the typical isotope shift between two adjacent isotopes of calcium
is about 160 - 200 MHz. The natural linewidth of the cooling transition is 34 MHz, and the Doppler shift for the
calcium atoms is 2.1 MHz/(m/s).
\begin{table}[hb]
\caption{\label{tab:isotopes}The abundances of the stable isotopes of calcium and the isotope shifts~\cite{15}
of the $4s^{2}$ $^{1}$S$_{0}$ - $4s4p$ $^{1}$P$_{1}$ transition. Also indicated is $^{41}$Ca. The shifts given
for $^{41}$Ca and $^{43}$Ca are for the 9/2 hyperfine component.}
\begin{center}
\begin{tabular}{c|c|c}

{\it Isotope} & {\it Natural abundance(\%)} & {\it Isotope Shift (MHz)}\\
\hline \hline
40 & $96.94$ & $0$\\
41 & $1 \cdot 10^{-12}$ & $166$\\
42 & $0.65$ & $393$\\
43 & $0.14$ & $554$\\
44 & $2.09$ & $774$\\
46 & $4 \cdot 10^{-3}$ & $1160$\\
48 & 0.19 & $1513$\\
\hline
\end{tabular}
\end{center}
\end{table}

Because the transverse Doppler broadening is limited the compression is isotope selective: when tuning the laser
frequency in between two adjacent isotopes there will be a force on both isotopes, opposite in sign. The result
is that the lighter isotope of the two will be pushed away from the beam axis while at the same time the
transverse velocity component of the heavier isotope can be reduced, resulting in an improved transmission
through the Zeeman slower. We have done Monte Carlo simulations to investigate this effect in more detail: a
detailed comparison with experimental results will be reported on in the near future.

The Zeeman slower is designed to decelerate atoms with initial velocities up to 1000 m/s (corresponding to 85\%
of the atoms) down to 50 m/s. The total length is 0.86 meter. The magnetic field has a maximum of 0.16 Tesla at
the entrance of the Zeeman slower, and decreases towards the exit. The laser beam counter-propagating the atoms
has a $\sigma^{+}$ polarization and a typical power of 20 mW. To avoid an extra velocity spread of the atoms,
the slowing process has to be terminated quickly, cf. reference ~\cite{87}. Therefore, the magnetic field has a
negative offset (as a result the slowing laser is detuned by -320 MHz) and is terminated with a strong positive
magnetic field. The laser beam is focused on the exit channels of the oven. As a result there is a gradient of
the laser intensity along the Zeeman slower, for which the shape of the magnetic field slope has been adjusted.
The coil generating the magnetic field is divided in 8 independently adjustable sections so that the field can
be optimized.

In the deflection chamber the desired isotope is deflected out of the slowed atom beam by a combination of two
counter-propagating red-detuned laser beams which cross the atom beam under an angle of 30$^{\circ}$. At this
point the Doppler broadening is sufficiently reduced so that direct isotope selection is possible, even when
probing the beam under 30 degrees. At this point the laser beam has a diameter of only 0.5 cm, the same detuning
as the trapping beams, and has a typical power of 5 mW. Only the selected isotope will be deflected so that it
can reach the trap, which is located 40 cm further downstream.

The atoms are trapped by a standard Magneto-Optical Trap. Typically 15 mW of laser power is split over 3 laser
beams which are retro-reflected to make the 6 required trapping beams. The magnetic quadrupole field is
generated by two coils in anti-Helmholtz configuration placed outside the vacuum (902 windings each, typical
current 3 A) resulting in a magnetic field gradient of 8 mT/cm. The fluorescence of the atoms in the trap is
measured by a photo-multiplier tube (PMT) in photon-counting mode. A lens system of 2.54 cm diameter is placed
inside the chamber at a focal distance of 3 cm from the trapped atoms to efficiently collect the fluorescence.
This lens system is adapted from reference \cite{13}. Outside the vacuum trapping chamber the light from the
trap is focused through a pinhole before detection by the PMT to reduce background light. For a large total
dynamic range of the detection system also a CCD camera and a photo-diode are used. The windows of the MOT have
been mounted on extension tubes and are anti-reflection coated. In order to further reduce the amount of
scattered light the inside of the vacuum chamber is coated black. The typical pressure in the MOT is $10^{-9}$
mbar which can rise to $10^{-8}$ mbar depending on the oven temperature.

The abundance of very rare isotopes can no longer be determined directly from the intensity of the fluorescence
from the trapped atoms if the average population of the trap falls below one atom. In this case we have to be
able to count the number of atoms arriving and leaving the trap over a certain period of time, and determine the
abundance from a comparison with the known abundance of another isotope. Experimental results on the detection
of the single atoms are presented and discussed in section \ref{sec:singleatoms}.

The present sensitivity of $^{41}$Ca detection by ATTA, as reported by Moore et al \cite{65}, is limited by the
background of $^{40}$Ca atoms present in the trap, and by the trapping time of the atoms. Even though the
$^{40}$Ca atoms are not trapped while detecting $^{41}$Ca, they contribute to the fluorescence signal from the
trap, obscuring the signal from the less abundant isotopes. It is therefore a key issue to further increase the
isotope selectivity. Besides the fluorescence background from these $^{40}$Ca atoms, there is the fluorescence
background due to scattering of the trapping laser beams from the windows and walls of the trapping chamber. For
any measurement of single atoms in a MOT this source of background has to be made as small as possible.

We have compared the isotope selectivity of three different trap loading schemes. As an indication of the
isotope selectivity of our system the amount of fluorescence due to the hot $^{40}$Ca atoms is measured while
trapping the calcium isotope $^{43}$Ca. The experimental results are presented in the next section.

\section{Results}\label{sec:results}
\begin{figure}[htb]
\includegraphics[width=\columnwidth]{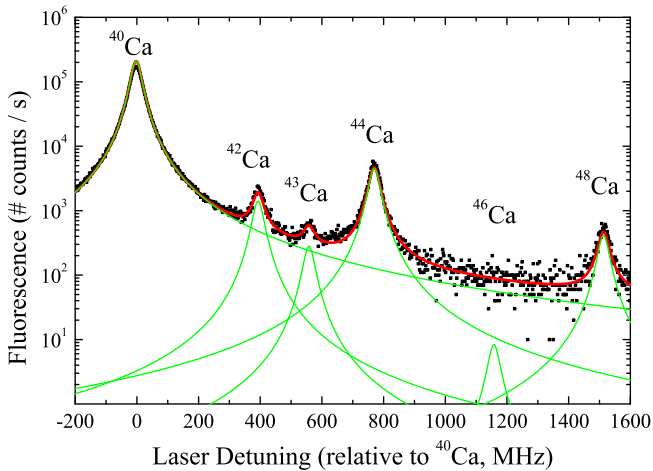}
\caption{\label{fig:figure2}The fluorescence of calcium isotopes in an atomic beam after a single excitation
step.}
\end{figure}
The starting point of the experiment is the atomic beam as it leaves the oven. The natural abundance of the
various isotopes in this beam can be seen in figure~\ref{fig:figure2}. In this figure the fluorescence of the
atomic beam coming from oven 2 is shown, excited only by the vertical trapping laser beam in the MOT. The
fluorescence is measured by a PMT under right angles with both the atom beam and the laser beam. The peaks have
been fitted with Lorentzian profiles, all having the same width (38 MHz) and an amplitude corresponding to the
natural abundance as given in table~\ref{tab:isotopes}. The fit curve is in good agreement with the measured
spectrum. The measured width of 38 MHz is slightly larger than the natural linewidth of 34 MHz: this corresponds
to a transverse velocity component of 1.9 m/s in the atomic beam. This figure gives an indication of the limited
maximum selectivity that can be reached in a single excitation step: the power of ATTA lies in the fact that the
same transition is excited very often.

\subsection{Comparing different loading schemes}\label{sec:comparing}
\begin{figure}[tb]
\includegraphics[width=\columnwidth]{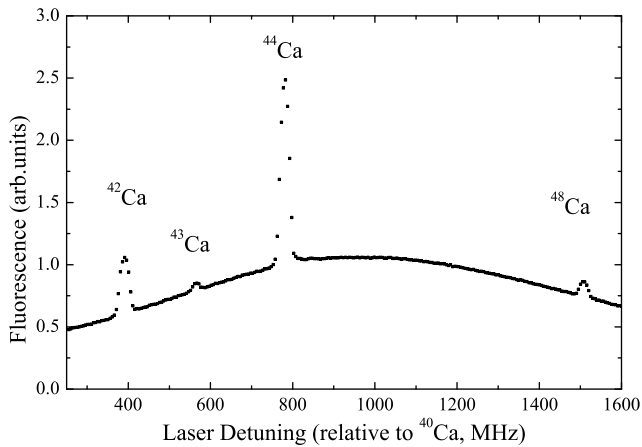}
\caption{\label{fig:figure3}Fluorescence in the trap loaded from oven 2. The background signal is caused by
$^{40}$Ca passing through the trap with its Maxwell-Boltzmann velocity distribution and excited by the
horizontal trapping laser beams.}
\end{figure}
\begin{figure}[tb]
\includegraphics[width=\columnwidth]{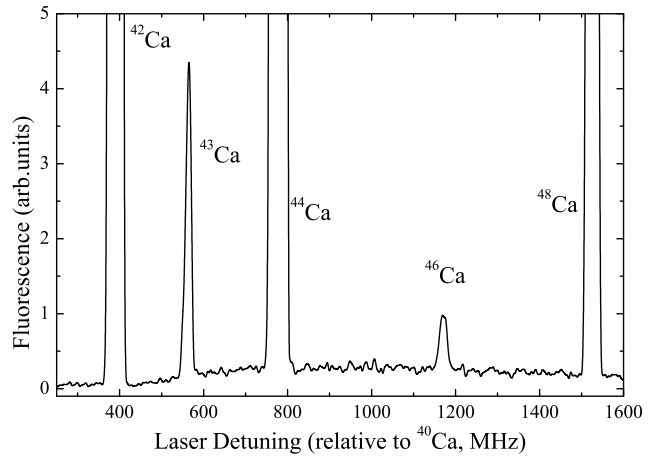}
\caption{\label{fig:figure4}Fluorescence in the trap loaded from a slowed but not deflected atom beam. This
spectrum was taken before the deflection stage was installed (cf. fig. \ref{fig:setup})}
\end{figure}
\begin{figure}[tb]
\includegraphics[width=\columnwidth]{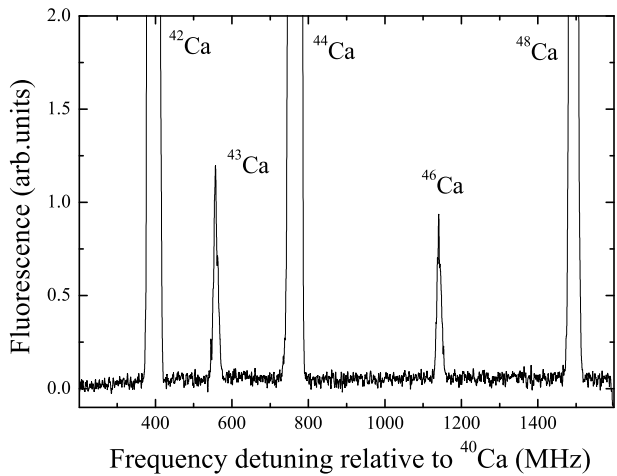}
\caption{\label{fig:figure5}Fluorescence in the trap loaded from a slowed and deflected beam.}
\end{figure}

In this section three different loading schemes are compared with respect to isotope selectivity. The simplest
configuration is to load the MOT directly from a thermal beam. Since the capture velocity of the trap is $\sim
50$ m/s only the low velocity tail of the Boltzmann distribution is trapped.  When the MOT is operated on one of
the heavier isotopes, the large amount of $^{40}$Ca limits the detection of these less abundant isotopes in the
trap. The horizontal trapping beams which intersect the atom beam at 45$^{\circ}$ can excite the fast $^{40}$Ca
atoms in the atom beam as the frequency of the trapping laser beams is scanned over the various isotopes. The
broad velocity distribution of these hot $^{40}$Ca atoms can be seen in figure~\ref{fig:figure3} as it dominates
the fluorescence at frequencies where for example trapped $^{46}$Ca atoms should be visible. To enable
comparison with figures \ref{fig:figure4} and \ref{fig:figure5}, the fluorescence is set to 1 at the resonance
frequency of $^{46}$Ca (1160 MHz). The hot atoms not only scatter the laser-light, thereby obscuring the signal
from the trapped atoms, but also shorten the trapping time of other isotopes by collisions with the trapped
atoms. The ratio between the fluorescence of the trapped $^{43}$Ca and the background $^{40}$Ca is $\sim 0.15$.

When the trap is loaded from a slowed atom beam the amount of trapped atoms increases because a larger fraction
of the atoms falls within the capture range of the trap. Figure~\ref{fig:figure4} shows the fluorescence
detected in the MOT, while simultaneously scanning the frequency of the trapping laser and the Zeeman slower
laser beams over the range of isotopes. This measurement was done before the deflection stage was installed. It
can be seen that the relative contribution of the hot $^{40}$Ca atoms is greatly reduced. Comparing
figure~\ref{fig:figure3} to figure~\ref{fig:figure4}, the ratio of for example the trapped $^{44}$Ca to the
$^{40}$Ca-background increases from $2.5$ to $5000$. The reduction of the $^{40}$Ca-background enables the
detection of trapped $^{46}$Ca atoms with a natural abundance of only 0.004 \%, which was not possible when
loading the trap directly from the thermal beam. The figure is normalized to the intensity of the $^{46}$Ca
peak. The ratio between the trapped $^{46}$Ca and the background $^{40}$Ca is $\sim 3$, for $^{43}$Ca the ratio
to the $^{40}$Ca background is $\sim 15$. The increase of isotope selectivity is the result of both the
increased fraction of slow atoms and the isotope selectivity of the Zeeman slower itself. In section
\ref{sec:simulations} we will look at this in some more detail.

\begin{figure}[tb]
\includegraphics[width=\columnwidth]{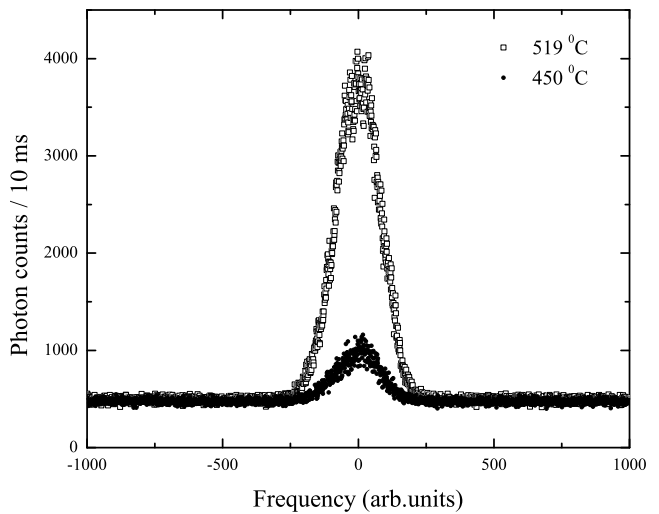}
\caption{\label{fig:peakback}Fluorescence intensity of $^{43}$Ca and background level intensity for two
different oven temperatures. The $^{43}$Ca peak intensity increases by a factor of 6.6, while the fluorescence
background level remains almost constant.}
\end{figure}

We find that the background signal from fast $^{40}$Ca atoms disappears almost completely when loading the trap
from a deflected, slow beam. In the measurement presented in figure~\ref{fig:figure5} the background in between
all the trapped isotopes is found to be mostly dependent on the intensity of the laser trapping beams, and only
slightly on the oven temperature (atom flux). This is further illustrated by figure~\ref{fig:peakback}. Here the
fluorescence from the trap is shown while scanning over the $^{43}$Ca trapping frequency for two different oven
temperatures. While the $^{43}$Ca peak intensity increases by a factor of 6.6 (from $\sim 500$ to $\sim 3300$
counts/10 ms), the average background level changes only from $475.9 \pm 0.7$ to $492.5 \pm 0.5$ counts/10 ms.
This insignificant background increase of less then 4 \%, corresponding to $16.6 \pm 0.6$, is due to extra
$^{40}$Ca atoms. The purely $^{40}$Ca dependent background component should also have increased by a factor of
6.6: therefore we can conclude that at 450 $^{\circ}$C the background contribution due to $^{40}$Ca is $2.5 \pm
0.1$ counts/10 ms. The $^{43}$Ca signal relative to the $^{40}$Ca background is therefore $\sim 200$. When
comparing this to a directly trapped thermal beam (figure \ref{fig:figure3}) the ratio of the $^{43}$Ca signal
to the $^{40}$Ca background is increased by a factor of $\sim 1300$. The rest of the background is due to laser
light scattered from the walls and the windows of the trapping chamber, and can be further reduced to 125 counts
/ 10 ms as shown by the data on single atoms, presented in section \ref{sec:singleatoms}. It is noted that the
measured $^{43}$Ca to $^{46}$Ca ratio changes, when comparing figure \ref{fig:figure4} to figure
\ref{fig:figure5}. This is discussed in section \ref{sec:fractionation}.

\subsection{Monte Carlo simulations}\label{sec:simulations}
\begin{figure}[tb]
\includegraphics[width=\columnwidth]{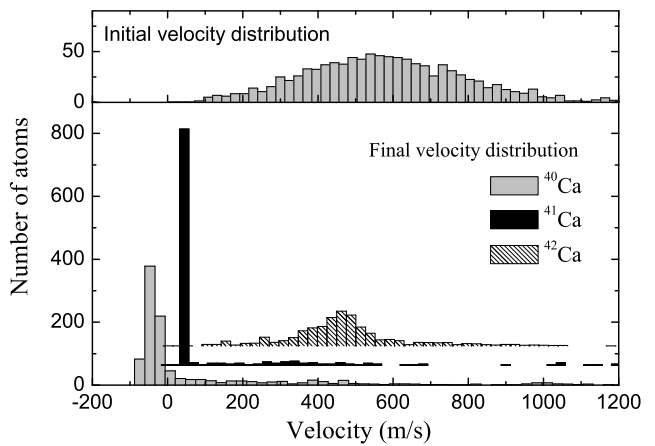}
\caption{\label{fig:zeemansl}A calculation of the final longitudinal velocity distribution (1000 atoms each) of
the isotopes $^{40}$Ca, $^{41}$Ca and $^{42}$Ca travelling through the Zeeman slower. The laser detuning is set
for $^{41}$Ca. An example of a Monte Carlo sampling of the initial velocity distribution upon entering the
Zeeman slower is shown in the top panel.}
\end{figure}
The isotope selectivity of the Zeeman slower can be understood on basis of the results of Monte Carlo
simulations. In the simulation results depicted in figure~\ref{fig:zeemansl}, the final velocity distributions
for $^{40}$Ca, $^{41}$Ca and $^{42}$Ca atoms were calculated for our Zeeman slower with the laser frequency
optimized for $^{41}$Ca. Most of the $^{41}$Ca atoms are decelerated to the desired velocity of about 50 m/s;
for both the other isotopes this is different. This can be understood as follows: for a given laser detuning all
the isotopes of calcium will be resonant in the Zeeman slower, but at different locations because of the isotope
shifts. These isotope shifts translate into different scattering rates in the Zeeman slower, according to
equation~\ref{eqn:scattering}. Therefore the final velocity of the different isotopes leaving the Zeeman slower
will be different. The magnetic field slope of the Zeeman slower is steeper at the end, where the average
velocity of the atoms is lower. If the laser is tuned to $^{41}$Ca atoms, then the $^{42}$Ca atoms will
initially also be decelerated, but the slope of the magnetic field is too steep: the change in velocity is not
enough to keep up with the change in magnetic field. Therefore this isotope will be lost from the slowing
process. On the other hand, a $^{40}$Ca atom is nicely slowed down until the end, but the slowing process
continues too long. In the simulation the far majority of these atoms undergo a reversal of their longitudinal
velocity. They will thus be lost even due to their small transversal velocity component and do not reach the
MOT.

As can be seen from figure~\ref{fig:figure4} a significantly lower background from $^{40}$Ca is detected in the
MOT for a trapping detuning smaller than about 450 MHz. With this detuning we probe $^{40}$Ca atoms with a
velocity range below 200 m/s. Since there is at the moment no collimation of the atom beam at the exit of the
Zeeman slower these atoms probably do not reach the trap due to the divergence of the atomic beam. The Zeeman
slower is designed to slow the desired isotope to 50 m/s; from the fact that we trap these isotopes we conclude
that for these isotopes the gain due to the larger number of slow atoms outweighs the loss due to the transverse
velocity spread. However, many more of these slow atoms could reach the trap if we implement a second
transversal cooling stage at the exit of the Zeeman slower.
\begin{figure}[tb]
\includegraphics[width=\columnwidth]{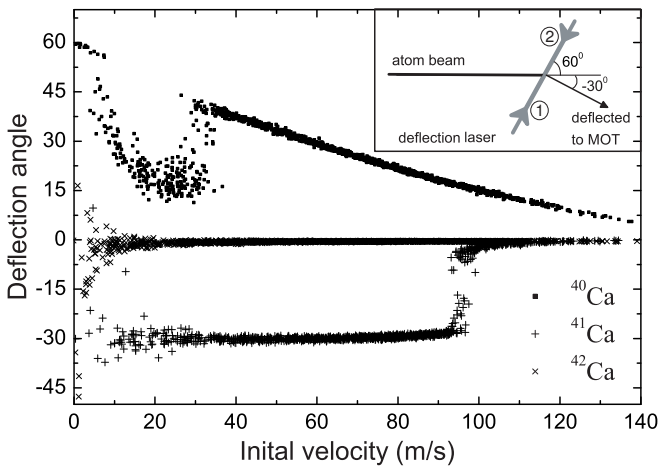}
\caption{\label{fig:simulation}A simulation of the large-angle atomic beam deflection for the different
isotopes. Shown is the deflection angle as a function of the initial velocity for three different isotopes:
$^{40}$Ca, $^{41}$Ca and $^{42}$Ca. The inset shows the geometry of the atomic beam and the deflection laser
beams. The laser detuning is 40 MHz red detuned with respect to the $^{41}$Ca resonance frequency. Further
details are given in the text.}
\end{figure}

To assess the functionality of the large-angle deflection by optical molasses we have performed Monte-Carlo
simulations. $^{40}$Ca, $^{41}$Ca and $^{42}$Ca atoms with a range of longitudinal velocities have been traced
through the deflection molasses. Absorption and emission probabilities are calculated along the way, and every
time a photon is absorbed and emitted the velocity of the atom is adjusted correspondingly. Stimulated emission
is also accounted for. Plotted in figure~\ref{fig:simulation} are the deflection angles as a function of the
initial velocity for $^{40}$Ca, $^{41}$Ca, and $^{42}$Ca atoms. In the inset the geometry of the atomic beam and
the deflection laser beams is shown. As in the experiment, the laser power is 5 mW distributed uniformly over a
beam diameter of 0.5 cm. The deflection laser is tuned 40 MHz below the resonance frequency of $^{41}$Ca. While
the $^{41}$Ca atoms are deflected well for velocities up to 90 m/s, the $^{40}$Ca are pushed away from the
atomic beam axis. This is the main reason that the combination of a Zeeman slower and the large-angle deflection
is so effective in selectively deflecting only one desired isotope out of the atom beam. Due to the large
additional detuning caused by the isotope shift, $^{42}$Ca is not affected at all. Similar simulations have been
done for different laser power, and it follows that both the isotope selectivity and the efficiency of the
deflection stage increase with increasing laser power.

For $^{40}$Ca we can analyze the deflection behavior in some more detail. Only the velocity component along the
axis of the molasses laser beams (called $v_{mol})$ is affected by the deflection molasses. The $^{40}$Ca atoms
are much closer to resonance to the molasses laser beam pushing them away from the trap (laser 2, cf. fig
\ref{fig:simulation}). The number of scattered photons depends on the scattering rate as given in
formula~\ref{eqn:scattering}, and on the time the atoms spend in the optical molasses. For the $^{40}$Ca atoms
at the given detuning this scattering rate is a Lorentzian with its maximum at $v_{mol} = 50$ m/s. This
corresponds to an atom with a longitudinal velocity (called $v_{long}$) of 100 m/s. During the time that the
atoms are in the optical molasses $v_{mol}$ will change due to the scattering. The atoms can be accelerated to a
final $v_{mol}$ which depends on the time spent in the molasses and the linewidth of the transition. The
deflection angle is determined by the ratio between $v_{mol}$ and $v_{long}$. The minimum in the deflection
angle for $^{40}$Ca atoms around 20 m/s is the region where $v_{mol}$ is limited by the time the atom spent in
the molasses. For $v_{long}<5$ m/s the deflection reaches the maximum possible deflection angle of 60$^{\circ}$,
i.e. parallel to the molasses laser beams. For $v_{long}>30$ m/s $v_{mol}$ reaches a maximum value limited by
the linewidth of the transition. The deflection angle is then just determined by the ratio of this maximum value
of $v_{mol}$ and the initial longitudinal velocity.

\subsection{Single atoms}\label{sec:singleatoms}
\begin{figure}[bt]
\includegraphics[width=\columnwidth]{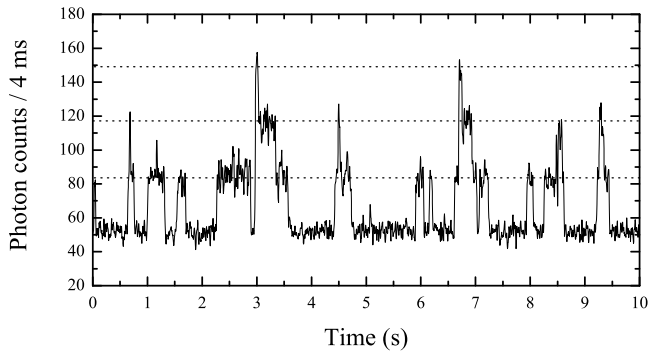}
\caption{\label{fig:singleatoms}Fluorescence of individual atoms detected in the MOT. The dotted lines indicate
the fluorescence level for 1, 2 and 3 atoms in the trap. The integration time of the photon-detector is 4 ms.}
\end{figure}
Single atoms can be detected in a magneto-optical trap because the trapped atoms are continuously scattering
photons from the trapping laser beams. Provided the amount of detected photons not originating from the trapped
atom is low enough, a single trapped atom can be detected as a temporal burst of light. In order to test the
sensitivity of the detection system, we have detected the fluorescence of trapped calcium atoms for low oven
temperatures. With decreasing oven temperature the atomic beam becomes less and less intense, down to a loading
rate of only a few atoms per minute. At some point the fluorescence from the trap displays discrete steps: we
can count the number of atoms in the trap. Shown in figure \ref{fig:singleatoms} is the fluorescence of single
$^{40}$Ca atoms detected in the trap for a period of 10 seconds. We have also detected single atoms from all
other stable calcium isotopes. During the time that an atom is trapped it can scatter photons at an estimated
rate (depending on laser detuning and intensity) of $1.3 \cdot 10{^6}$ s$^{-1}$. We detect $35 \pm 3$ photons
per 4 ms per atom, which results in a total photon detection efficiency of 0.7 \%. The background level is 50
photons per 4 ms. This should not be confused with the noise level, which is limited by statistics. Trapping the
atoms longer enables us to use a larger integration time which reduces the statistical error on the signal.
During the measurement of figure \ref{fig:singleatoms} the repump laser was used: we concluded from trap decay
time measurements that the average trapping time improved from $\sim$ 20 ms to $\sim$ 200 ms. It is important to
note that the background contribution due to hot $^{40}$Ca atoms such as reported in section
\ref{sec:comparing}, even at high oven temperatures, is much lower then the fluorescence signal of a single
trapped atom.

\subsection{Isotope fractionation}\label{sec:fractionation}
The final issue we will address in this article is that of isotope fractionation effects in the experiment. The
maximum fluorescence of the various trapped isotopes in figure~\ref{fig:figure5} is compared to literature
values~\cite{248} in table~\ref{tab:abundance}. The absolute amount of $^{40}$Ca could not be measured
accurately in the same experiment due to the high intensity of the fluorescence: therefore the relative
abundance of the heavier isotopes was compared to $^{44}$Ca. The abundance of $^{44}$Ca was set to the
literature value of $2.09$.
\begin{table}[h]
\caption{\label{tab:abundance}The measured abundance of the stable calcium isotopes compared to literature
values from~\cite{248}. $^{44}$Ca is taken as a reference point, and for $^{43}$Ca two values are shown for two
different laser powers (50 mW / 70 mW).}
\begin{center}
\begin{tabular}{l|l|l}
{\it Isotope} & {\it Measured(\%)} & {\it Literature(\%)}\\
\hline \hline
40 & $-$ & $96.94(16)$\\
42 & $0.68$ & $0.65(2)$\\
43 & $0.006 / 0.06$ & $0.14(1)$\\
44 & $[2.09]$ & $2.09(11)$\\
46 & $0.005$ & $0.004(3)$\\
48 & $0.15$ & $0.19(2)$\\
\hline
\end{tabular}
\end{center}
\end{table}

For all isotopes except $^{43}$Ca the agreement is reasonable. In the case of $^{43}$Ca we detect an amount of
fluorescence which would indicate an abundance of only $0.006 \%$ for the typically used laser power of $\sim$
40 mW. This corresponds to only $\sim$ 4 \% of the literature value. This large discrepancy is due to
differences in the cooling and trapping efficiency of the odd isotopes of calcium: the odd isotopes $^{41}$Ca
and $^{43}$Ca have a nuclear spin of I=7/2. The resulting magnetic substructure of the ground and the excited
state influences the Doppler cooling force. Comparable observations have been reported for the case if the odd
strontium isotopes~\cite{236}. We have adapted the theoretical model developed for strontium and have solved the
generalized optical Bloch equations for the odd calcium isotopes taking into account all the hyperfine states
and their magnetic sub-states. We find that the maximum deceleration force on the odd isotopes for a
one-dimensional $\sigma^{+}$-$\sigma^{-}$ optical molasses configuration is only half as strong as that on the
even isotopes. This indicates that the number of photons emitted per trapped atom in a certain period of time is
only half that of the even isotopes. This alone cannot explain the reduction of a factor of 25 in the detection
of $^{43}$Ca. Since the cooling force is weaker this effect could also limit the efficiency of the deflection
and slowing of the odd isotopes. In order to investigate this issue, we have done a recent experiment where we
find that the measured $^{43}$Ca abundance depends significantly on the laser-power. Increasing the total
laser-power from 50 mW to 70 mW the ratio between $^{43}$Ca and $^{46}$Ca increases from 1.2 to 15. This ratio
can be more accurately measured than the relative abundances, due to non-linear effects in the detection system
for the very high count rates when measuring $^{44}$Ca. A ratio of 15 corresponds to a measured abundance for
$^{43}$Ca of 0.06 \%: this is roughly half of the literature value. This observation indicates clearly that the
efficiency of the deflection and the Zeeman slower is different for the odd and even isotopes, and depends
critically on the laser power available. Thus for the planned investigations on $^{41}$Ca, the isotope $^{43}$Ca
should be chosen as reference, due to its expected similar behavior in the experimental scheme.

\section{Conclusion}
We have demonstrated the trapping of all stable isotopes of calcium involving cooling in a Zeeman slower and
deflection of the cooled beam by 30$^{\circ}$ in a tilted optical molasses. A corresponding experimental system
has been built with a sensitivity that enables us to detect even single trapped atoms of all these isotopes. To
suppress the interfering $^{40}$Ca fluorescence background when detecting the less abundant isotopes the
deflection of the slowed loading beam over 30$^{\circ}$ was found to be very effective. The background due to
$^{40}$Ca decreased by three orders of magnitude compared to direct loading from a thermal atomic beam. We find
that finally the background signal for single atom detection of the less abundant isotopes is no longer limited
by the $^{40}$Ca atoms but dominated by laser stray light. This enables single atom detection of all stable
isotopes. Improvements will be made to the setup: increasing the laser power will increase the efficiency and
further improve the performance and the isotope selectivity of the individual components: transversal beam
compression, Zeeman slower and deflection stage. Adding a second isotope selective compression stage directly
after the Zeeman slower is another potential improvement. With these changes the system will be ready to start
searching for $^{41}$Ca atoms at the natural abundance level.

\begin{acknowledgments} The authors would like to thank the KVI technical staff (J. Mulder and J. Sa)
for their support, C. van Ditzhuijzen for her work on the Monte-Carlo simulations, J. W. Dunn for providing the
code to take into account the multilevel structure of $^{43}$Ca and Toptica Photonics for prompt assistance.
This project (00PR1887) is part of the research program of the Stichting voor Fundamenteel Onderzoek der Materie
(FOM) which is supported by the Nederlandse Organisatie voor Wetenschappelijk Onderzoek (NWO). Additional
support is from the EU contract NIPNET (HPRI-CT-2001-50034).
\end{acknowledgments}
\bibliographystyle{apsrev}
\bibliography{references}

\begin{thebibliography}{23}
\expandafter\ifx\csname natexlab\endcsname\relax\def\natexlab#1{#1}\fi
\expandafter\ifx\csname bibnamefont\endcsname\relax
  \def\bibnamefont#1{#1}\fi
\expandafter\ifx\csname bibfnamefont\endcsname\relax
  \def\bibfnamefont#1{#1}\fi
\expandafter\ifx\csname citenamefont\endcsname\relax
  \def\citenamefont#1{#1}\fi
\expandafter\ifx\csname url\endcsname\relax
  \def\url#1{\texttt{#1}}\fi
\expandafter\ifx\csname urlprefix\endcsname\relax\def\urlprefix{URL }\fi
\providecommand{\bibinfo}[2]{#2}
\providecommand{\eprint}[2][]{\url{#2}}

\bibitem[{\citenamefont{Lu and Wendt}(2003)}]{60}
\bibinfo{author}{\bibfnamefont{Z.~T.} \bibnamefont{Lu}} \bibnamefont{and}
  \bibinfo{author}{\bibfnamefont{K.~D.~A.} \bibnamefont{Wendt}},
  \bibinfo{journal}{Rev. Sci. Instrum.} \textbf{\bibinfo{volume}{74}},
  \bibinfo{pages}{1169} (\bibinfo{year}{2003}).

\bibitem[{\citenamefont{Chen et~al.}(1999)\citenamefont{Chen, Li, Bailey,
  O'Connor, Young, and Lu}}]{223}
\bibinfo{author}{\bibfnamefont{C.~Y.} \bibnamefont{Chen}},
  \bibinfo{author}{\bibfnamefont{Y.~M.} \bibnamefont{Li}},
  \bibinfo{author}{\bibfnamefont{K.}~\bibnamefont{Bailey}},
  \bibinfo{author}{\bibfnamefont{T.~P.} \bibnamefont{O'Connor}},
  \bibinfo{author}{\bibfnamefont{L.}~\bibnamefont{Young}}, \bibnamefont{and}
  \bibinfo{author}{\bibfnamefont{Z.~T.} \bibnamefont{Lu}},
  \bibinfo{journal}{Science} \textbf{\bibinfo{volume}{286}},
  \bibinfo{pages}{1139} (\bibinfo{year}{1999}).

\bibitem[{\citenamefont{Bailey et~al.}(2000)\citenamefont{Bailey, Chen, Du, Li,
  Lu, O'Connor, and Young}}]{213}
\bibinfo{author}{\bibfnamefont{K.}~\bibnamefont{Bailey}},
  \bibinfo{author}{\bibfnamefont{C.~Y.} \bibnamefont{Chen}},
  \bibinfo{author}{\bibfnamefont{X.}~\bibnamefont{Du}},
  \bibinfo{author}{\bibfnamefont{Y.~M.} \bibnamefont{Li}},
  \bibinfo{author}{\bibfnamefont{Z.~T.} \bibnamefont{Lu}},
  \bibinfo{author}{\bibfnamefont{T.~P.} \bibnamefont{O'Connor}},
  \bibnamefont{and} \bibinfo{author}{\bibfnamefont{L.}~\bibnamefont{Young}},
  \bibinfo{journal}{Nucl. Instrum. Methods B} \textbf{\bibinfo{volume}{172}},
  \bibinfo{pages}{224} (\bibinfo{year}{2000}).

\bibitem[{\citenamefont{Sturchio et~al.}(2004)\citenamefont{Sturchio, Du,
  Purtschert, Lehmann, Sultan, Patterson, Lu, Muller, Bigler, Bailey
  et~al.}}]{215}
\bibinfo{author}{\bibfnamefont{N.~C.} \bibnamefont{Sturchio}},
  \bibinfo{author}{\bibfnamefont{X.}~\bibnamefont{Du}},
  \bibinfo{author}{\bibfnamefont{R.}~\bibnamefont{Purtschert}},
  \bibinfo{author}{\bibfnamefont{B.~E.} \bibnamefont{Lehmann}},
  \bibinfo{author}{\bibfnamefont{M.}~\bibnamefont{Sultan}},
  \bibinfo{author}{\bibfnamefont{L.~J.} \bibnamefont{Patterson}},
  \bibinfo{author}{\bibfnamefont{Z.~T.} \bibnamefont{Lu}},
  \bibinfo{author}{\bibfnamefont{P.}~\bibnamefont{Muller}},
  \bibinfo{author}{\bibfnamefont{T.}~\bibnamefont{Bigler}},
  \bibinfo{author}{\bibfnamefont{K.}~\bibnamefont{Bailey}},
  \bibnamefont{et~al.}, \bibinfo{journal}{Geophys. Res. Lett.}
  \textbf{\bibinfo{volume}{31}}, \bibinfo{pages}{L05503}
  (\bibinfo{year}{2004}).

\bibitem[{\citenamefont{Henning et~al.}(1987)\citenamefont{Henning, Bell,
  Billquist, Glagola, Kutschera, Liu, Lucas, Paul, Rehm, and Yntema}}]{111}
\bibinfo{author}{\bibfnamefont{W.}~\bibnamefont{Henning}},
  \bibinfo{author}{\bibfnamefont{W.~A.} \bibnamefont{Bell}},
  \bibinfo{author}{\bibfnamefont{P.~J.} \bibnamefont{Billquist}},
  \bibinfo{author}{\bibfnamefont{B.~G.} \bibnamefont{Glagola}},
  \bibinfo{author}{\bibfnamefont{W.}~\bibnamefont{Kutschera}},
  \bibinfo{author}{\bibfnamefont{Z.}~\bibnamefont{Liu}},
  \bibinfo{author}{\bibfnamefont{H.~F.} \bibnamefont{Lucas}},
  \bibinfo{author}{\bibfnamefont{M.}~\bibnamefont{Paul}},
  \bibinfo{author}{\bibfnamefont{K.~E.} \bibnamefont{Rehm}}, \bibnamefont{and}
  \bibinfo{author}{\bibfnamefont{J.~L.} \bibnamefont{Yntema}},
  \bibinfo{journal}{Science} \textbf{\bibinfo{volume}{236}},
  \bibinfo{pages}{725} (\bibinfo{year}{1987}).

\bibitem[{\citenamefont{Taylor}(1987)}]{110}
\bibinfo{author}{\bibfnamefont{R.~E.} \bibnamefont{Taylor}},
  \bibinfo{journal}{Anal. Chem.} \textbf{\bibinfo{volume}{59}},
  \bibinfo{pages}{A317} (\bibinfo{year}{1987}).

\bibitem[{\citenamefont{Freeman et~al.}(2000)\citenamefont{Freeman, Beck,
  Bierman, Caffee, Heaney, Holloway, Marcus, Southon, and Vogel}}]{109}
\bibinfo{author}{\bibfnamefont{S.~P. H.~T.} \bibnamefont{Freeman}},
  \bibinfo{author}{\bibfnamefont{B.}~\bibnamefont{Beck}},
  \bibinfo{author}{\bibfnamefont{J.~M.} \bibnamefont{Bierman}},
  \bibinfo{author}{\bibfnamefont{M.~W.} \bibnamefont{Caffee}},
  \bibinfo{author}{\bibfnamefont{R.~P.} \bibnamefont{Heaney}},
  \bibinfo{author}{\bibfnamefont{L.}~\bibnamefont{Holloway}},
  \bibinfo{author}{\bibfnamefont{R.}~\bibnamefont{Marcus}},
  \bibinfo{author}{\bibfnamefont{J.~R.} \bibnamefont{Southon}},
  \bibnamefont{and} \bibinfo{author}{\bibfnamefont{J.~S.} \bibnamefont{Vogel}},
  \bibinfo{journal}{Nucl. Instrum. Methods B} \textbf{\bibinfo{volume}{172}},
  \bibinfo{pages}{930} (\bibinfo{year}{2000}).

\bibitem[{\citenamefont{Du et~al.}(2003)\citenamefont{Du, Purtschert, Bailey,
  Lehmann, Lorenzo, Lu, Mueller, O'Connor, Sturchio, and Young}}]{217}
\bibinfo{author}{\bibfnamefont{X.}~\bibnamefont{Du}},
  \bibinfo{author}{\bibfnamefont{R.}~\bibnamefont{Purtschert}},
  \bibinfo{author}{\bibfnamefont{K.}~\bibnamefont{Bailey}},
  \bibinfo{author}{\bibfnamefont{B.~E.} \bibnamefont{Lehmann}},
  \bibinfo{author}{\bibfnamefont{R.}~\bibnamefont{Lorenzo}},
  \bibinfo{author}{\bibfnamefont{Z.~T.} \bibnamefont{Lu}},
  \bibinfo{author}{\bibfnamefont{P.}~\bibnamefont{Mueller}},
  \bibinfo{author}{\bibfnamefont{T.~P.} \bibnamefont{O'Connor}},
  \bibinfo{author}{\bibfnamefont{N.~C.} \bibnamefont{Sturchio}},
  \bibnamefont{and} \bibinfo{author}{\bibfnamefont{L.}~\bibnamefont{Young}},
  \bibinfo{journal}{Geophys. Res. Lett.} \textbf{\bibinfo{volume}{30}}
  (\bibinfo{year}{2003}).

\bibitem[{\citenamefont{M\"uller et~al.}(2001)\citenamefont{M\"uller, Bushaw,
  Blaum, Diel, Geppert, N\"ahler, Trautmann, N\"ortersh\"auser, and
  Wendt}}]{312}
\bibinfo{author}{\bibfnamefont{P.}~\bibnamefont{M\"uller}},
  \bibinfo{author}{\bibfnamefont{B.~A.} \bibnamefont{Bushaw}},
  \bibinfo{author}{\bibfnamefont{K.}~\bibnamefont{Blaum}},
  \bibinfo{author}{\bibfnamefont{S.}~\bibnamefont{Diel}},
  \bibinfo{author}{\bibfnamefont{C.}~\bibnamefont{Geppert}},
  \bibinfo{author}{\bibfnamefont{A.}~\bibnamefont{N\"ahler}},
  \bibinfo{author}{\bibfnamefont{N.}~\bibnamefont{Trautmann}},
  \bibinfo{author}{\bibfnamefont{W.}~\bibnamefont{N\"ortersh\"auser}},
  \bibnamefont{and} \bibinfo{author}{\bibfnamefont{K.}~\bibnamefont{Wendt}},
  \bibinfo{journal}{Fresenius Journal of Analytical Chemistry}
  \textbf{\bibinfo{volume}{370}}, \bibinfo{pages}{508} (\bibinfo{year}{2001}).

\bibitem[{\citenamefont{Wendt et~al.}(2003)\citenamefont{Wendt, Blaum, Geppert,
  Horn, Passler, Trautmann, and Bushaw}}]{313}
\bibinfo{author}{\bibfnamefont{K.~D.~A.} \bibnamefont{Wendt}},
  \bibinfo{author}{\bibfnamefont{K.}~\bibnamefont{Blaum}},
  \bibinfo{author}{\bibfnamefont{C.}~\bibnamefont{Geppert}},
  \bibinfo{author}{\bibfnamefont{R.}~\bibnamefont{Horn}},
  \bibinfo{author}{\bibfnamefont{G.}~\bibnamefont{Passler}},
  \bibinfo{author}{\bibfnamefont{N.}~\bibnamefont{Trautmann}},
  \bibnamefont{and} \bibinfo{author}{\bibfnamefont{B.~A.}
  \bibnamefont{Bushaw}}, \bibinfo{journal}{Nucl. Instrum. Methods B}
  \textbf{\bibinfo{volume}{204}}, \bibinfo{pages}{325} (\bibinfo{year}{2003}).

\bibitem[{\citenamefont{Freeman et~al.}(2001)\citenamefont{Freeman, Wendt,
  Mueller, and Geppert}}]{315}
\bibinfo{author}{\bibfnamefont{S.}~\bibnamefont{Freeman}},
  \bibinfo{author}{\bibfnamefont{K.}~\bibnamefont{Wendt}},
  \bibinfo{author}{\bibfnamefont{P.}~\bibnamefont{Mueller}}, \bibnamefont{and}
  \bibinfo{author}{\bibfnamefont{C.}~\bibnamefont{Geppert}},
  \bibinfo{journal}{J. Bone Miner. Res.} \textbf{\bibinfo{volume}{16}},
  \bibinfo{pages}{S346} (\bibinfo{year}{2001}).

\bibitem[{\citenamefont{Moore et~al.}(2004)\citenamefont{Moore, Bailey, Greene,
  Lu, M\"uller, O'Connor, Geppert, Wendt, and Young}}]{65}
\bibinfo{author}{\bibfnamefont{I.~D.} \bibnamefont{Moore}},
  \bibinfo{author}{\bibfnamefont{K.}~\bibnamefont{Bailey}},
  \bibinfo{author}{\bibfnamefont{J.}~\bibnamefont{Greene}},
  \bibinfo{author}{\bibfnamefont{Z.~T.} \bibnamefont{Lu}},
  \bibinfo{author}{\bibfnamefont{P.}~\bibnamefont{M\"uller}},
  \bibinfo{author}{\bibfnamefont{T.~P.} \bibnamefont{O'Connor}},
  \bibinfo{author}{\bibfnamefont{C.}~\bibnamefont{Geppert}},
  \bibinfo{author}{\bibfnamefont{K.~D.~A.} \bibnamefont{Wendt}},
  \bibnamefont{and} \bibinfo{author}{\bibfnamefont{L.}~\bibnamefont{Young}},
  \bibinfo{journal}{Phys. Rev. Lett.} \textbf{\bibinfo{volume}{92}},
  \bibinfo{pages}{153002} (\bibinfo{year}{2004}).

\bibitem[{\citenamefont{Raab et~al.}(1987)\citenamefont{Raab, Prentiss, Cable,
  Chu, and Pritchard}}]{103}
\bibinfo{author}{\bibfnamefont{E.~L.} \bibnamefont{Raab}},
  \bibinfo{author}{\bibfnamefont{M.}~\bibnamefont{Prentiss}},
  \bibinfo{author}{\bibfnamefont{A.}~\bibnamefont{Cable}},
  \bibinfo{author}{\bibfnamefont{S.}~\bibnamefont{Chu}}, \bibnamefont{and}
  \bibinfo{author}{\bibfnamefont{D.~E.} \bibnamefont{Pritchard}},
  \bibinfo{journal}{Phys. Rev. Lett.} \textbf{\bibinfo{volume}{59}},
  \bibinfo{pages}{2631} (\bibinfo{year}{1987}).

\bibitem[{\citenamefont{Balykin et~al.}(1985)\citenamefont{Balykin, Letokhov,
  Minogin, Rozhdestvensky, and Sidorov}}]{249}
\bibinfo{author}{\bibfnamefont{V.~I.} \bibnamefont{Balykin}},
  \bibinfo{author}{\bibfnamefont{V.~S.} \bibnamefont{Letokhov}},
  \bibinfo{author}{\bibfnamefont{V.~G.} \bibnamefont{Minogin}},
  \bibinfo{author}{\bibfnamefont{Y.~V.} \bibnamefont{Rozhdestvensky}},
  \bibnamefont{and} \bibinfo{author}{\bibfnamefont{A.~I.}
  \bibnamefont{Sidorov}}, \bibinfo{journal}{J. Opt. Soc. Am. B}
  \textbf{\bibinfo{volume}{2}}, \bibinfo{pages}{1776} (\bibinfo{year}{1985}).

\bibitem[{\citenamefont{Phillips and Metcalf}(1982)}]{246}
\bibinfo{author}{\bibfnamefont{W.~D.} \bibnamefont{Phillips}} \bibnamefont{and}
  \bibinfo{author}{\bibfnamefont{H.}~\bibnamefont{Metcalf}},
  \bibinfo{journal}{Phys. Rev. Lett.} \textbf{\bibinfo{volume}{48}},
  \bibinfo{pages}{596} (\bibinfo{year}{1982}).

\bibitem[{\citenamefont{Wieman and H\"ansch}(1976)}]{4}
\bibinfo{author}{\bibfnamefont{C.}~\bibnamefont{Wieman}} \bibnamefont{and}
  \bibinfo{author}{\bibfnamefont{T.~W.} \bibnamefont{H\"ansch}},
  \bibinfo{journal}{Phys. Rev. Lett.} \textbf{\bibinfo{volume}{36}},
  \bibinfo{pages}{1170} (\bibinfo{year}{1976}).

\bibitem[{\citenamefont{Oates et~al.}(1999)\citenamefont{Oates, Bondu, Fox, and
  Hollberg}}]{54}
\bibinfo{author}{\bibfnamefont{C.~W.} \bibnamefont{Oates}},
  \bibinfo{author}{\bibfnamefont{F.}~\bibnamefont{Bondu}},
  \bibinfo{author}{\bibfnamefont{R.~W.} \bibnamefont{Fox}}, \bibnamefont{and}
  \bibinfo{author}{\bibfnamefont{L.}~\bibnamefont{Hollberg}},
  \bibinfo{journal}{European Physical Journal D} \textbf{\bibinfo{volume}{7}},
  \bibinfo{pages}{449} (\bibinfo{year}{1999}).

\bibitem[{\citenamefont{Metcalf and v.~d. Straten}(1999)}]{205}
\bibinfo{author}{\bibfnamefont{H.}~\bibnamefont{Metcalf}} \bibnamefont{and}
  \bibinfo{author}{\bibfnamefont{P.}~\bibnamefont{v.~d. Straten}},
  \emph{\bibinfo{title}{Laser Cooling and Trapping}}
  (\bibinfo{publisher}{Springer New York}, \bibinfo{year}{1999}).

\bibitem[{\citenamefont{N\"ortersh\"auser
  et~al.}(1998)\citenamefont{N\"ortersh\"auser, Trautmann, Wendt, and
  Bushaw}}]{15}
\bibinfo{author}{\bibfnamefont{W.}~\bibnamefont{N\"ortersh\"auser}},
  \bibinfo{author}{\bibfnamefont{N.}~\bibnamefont{Trautmann}},
  \bibinfo{author}{\bibfnamefont{K.}~\bibnamefont{Wendt}}, \bibnamefont{and}
  \bibinfo{author}{\bibfnamefont{B.~A.} \bibnamefont{Bushaw}},
  \bibinfo{journal}{Spectrochim. Acta, Part B} \textbf{\bibinfo{volume}{53}},
  \bibinfo{pages}{709} (\bibinfo{year}{1998}).

\bibitem[{\citenamefont{Lison et~al.}(2000)\citenamefont{Lison, Schuh,
  Haubrich, and Meschede}}]{87}
\bibinfo{author}{\bibfnamefont{F.}~\bibnamefont{Lison}},
  \bibinfo{author}{\bibfnamefont{P.}~\bibnamefont{Schuh}},
  \bibinfo{author}{\bibfnamefont{D.}~\bibnamefont{Haubrich}}, \bibnamefont{and}
  \bibinfo{author}{\bibfnamefont{D.}~\bibnamefont{Meschede}},
  \bibinfo{journal}{Phys. Rev. A} \textbf{\bibinfo{volume}{61}},
  \bibinfo{pages}{013405} (\bibinfo{year}{2000}).

\bibitem[{\citenamefont{Alt}(2002)}]{13}
\bibinfo{author}{\bibfnamefont{W.}~\bibnamefont{Alt}}, \bibinfo{journal}{Optik}
  \textbf{\bibinfo{volume}{113}}, \bibinfo{pages}{142} (\bibinfo{year}{2002}).

\bibitem[{\citenamefont{Rosman and Taylor}(1998)}]{248}
\bibinfo{author}{\bibfnamefont{K.~J.~R.} \bibnamefont{Rosman}}
  \bibnamefont{and} \bibinfo{author}{\bibfnamefont{P.~D.~P.}
  \bibnamefont{Taylor}}, \bibinfo{journal}{Pure Appl. Chem.}
  \textbf{\bibinfo{volume}{70}}, \bibinfo{pages}{217} (\bibinfo{year}{1998}).

\bibitem[{\citenamefont{Xu et~al.}(2003)\citenamefont{Xu, Loftus, Dunn, Greene,
  Hall, Gallagher, and Ye}}]{236}
\bibinfo{author}{\bibfnamefont{X.}~\bibnamefont{Xu}},
  \bibinfo{author}{\bibfnamefont{T.~H.} \bibnamefont{Loftus}},
  \bibinfo{author}{\bibfnamefont{J.~W.} \bibnamefont{Dunn}},
  \bibinfo{author}{\bibfnamefont{C.~H.} \bibnamefont{Greene}},
  \bibinfo{author}{\bibfnamefont{J.~L.} \bibnamefont{Hall}},
  \bibinfo{author}{\bibfnamefont{A.}~\bibnamefont{Gallagher}},
  \bibnamefont{and} \bibinfo{author}{\bibfnamefont{J.}~\bibnamefont{Ye}},
  \bibinfo{journal}{Phys. Rev. Lett.} \textbf{\bibinfo{volume}{90}},
  \bibinfo{pages}{193002} (\bibinfo{year}{2003}).

\end{thebibliography}
\end{document}